\documentclass[a4paper, 11pt, english]{IEEEtran}

\usepackage[T1]{fontenc}
\usepackage[english]{babel}
\usepackage{graphicx}
\usepackage{gensymb} % \degree
\usepackage{lettrine} % For dropcap
\usepackage{subcaption} % for adding subcaptions to a figure made out of several picture
\usepackage{floatrow} %  for a better control of the pictures position (\begin{figure}[H])
\usepackage{comment} 
\usepackage{hyperref}
\usepackage{cite} % for commenting sections of text
\usepackage[resetlabels,labeled]{multibib}
\usepackage[utf8]{inputenc}
\usepackage[table]{xcolor}
\usepackage{authblk}

\begin{document}

\setlength{\textfloatsep}{1\baselineskip plus 0.2\baselineskip minus 0.5\baselineskip}

%_________________________________________
\title{Plasma characterization and confirmation of helicon mode  of IShTAR plasma source }

\author[1]{ Foisal B. T. Siddiki}
\author[2]{I. Shesterikov}
\author[2]{J.-M. Noterdaeme$^{1,}$}
\author[1]{K. Crombé}
\affil[1]{Department of Applied Physics, Ghent University, Ghent, Belgium}
\affil[2]{Max Planck Institute for Plasma Physics, Garching, Germany}

\maketitle
\begin{abstract}

IShTAR (Ion cyclotron Sheath Test ARrangement) is a dedicated test facility to investigate the interaction of ICRF wave and plasma at the Max-Planck Institute for Plasma Physics in Garching, Germany. Plasma is provided by a plasma source (length= 0.1m, diameter = 0.4m) and it is responsible to create the necessary plasma environment in front of the ICRF antenna. The source is equipped with a helical antenna and designed to reach the mode of discharge. Previously there was no confirmation that this source can reach helicon mode of discharge. However, it is important for such a source to reach the helicon mode of discharge to work with its full capacity and to produce high-density plasma. To achieve helicon mode of discharge for this source a detailed study was carried out. Characterization of plasma parameters (i.e. electron temperature, density, and plasma potential) followed by the implementation of a global model of particle and power balance has been made. The existence of helicon mode is confirmed at lower magnetic fields (B <35 mT). To get the highest possible plasma density, the optimization of the plasma source is presented to reach helicon mode at the highest available magnetic field.

\end{abstract}
\smallskip
%\noindent \textbf{Keywords:} Helicon Plasma, Plasma source, 
\section{Introduction}
\lettrine[lines=2, findent=3pt, nindent=0pt] 
{H}{elicon} discharges considered as most efficient in terms of plasma production among other RF or DC discharges. In a helicon discharge electron densities can reach as high as $10^{-20} m^{-3}$ at a low temperature \cite{boswell1984very}\cite{buttenschon2018high}. Power absorb in a helicon discharge through helicon wave, which is a whistler wave in a bounded plasma \cite{chen1991plasma}. Although efficient plasma production mechanism in a helicon discharge by helicon waves poorly understood \cite{chen2015helicon}, they are being extensively used in many applications including material processing \cite{chen2001design}\cite{goulding2017progress}, for plasma production in toroidal devices\cite{tripathi2001normal}\cite{loewenhardt1991plasma}, in plasma propulsion \cite{diaz2001overview}\cite{winglee2007simulation}, negative ion production\cite{hayashi1998measurements}, thin-film deposition\cite{kim1997characterization}, plasma material interaction\cite{blackwell2012design} as well as widely used in research laboratories\cite{bohlin2014vineta}. Ion cyclotron Sheath Test Arrangement (IShTAR) is such a research laboratory where helicon type plasma source is being used \cite{crombe2016ishtar}. The main objective of this research facility is to understand spurious effects that degrade the overall performance of the ICRF heating operation in tokamak and to optimize the ICRF antenna \cite{crombe2016ishtar}\cite{crombe2018test}. To conduct this research it's important to mimic the tokamak edge plasma environment in-front of the testbed ICRF antenna and the IShTAR plasma source is solely responsible to create this edge plasma environment. Though this plasma source designed as a helicon source but previously there was no confirmation that this source can reach helicon mode of discharge. However, it is necessary for such a source which is designed as a helicon source to reach the helicon mode of discharge to realize its full potential for producing high dense plasma.  It is important for the experiment as well to reach the helicon mode of discharge to maintain fusion relevant plasma condition in front of the ICRF antenna.\\
\begin{figure*}[htb]
        \centerline{\includegraphics[width=.8\textwidth]{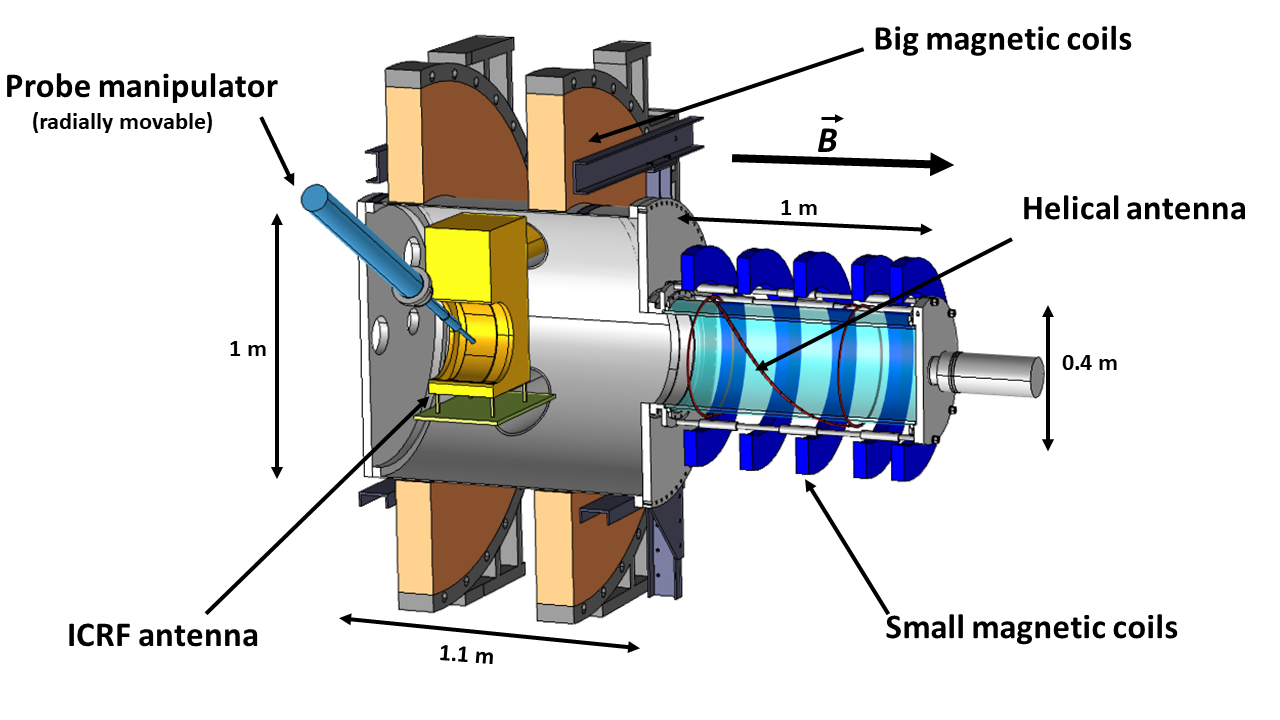} }
        \caption[]{Schematic of the axial cross-section of IShTAR test facility with all its main components \cite{usoltceva2019advancements}}
        \label{fig:cross-section}
 \end{figure*}
The characteristic feature of a helicon source is that it exhibits three modes of power coupling. When the strength of RF power or magnetic field varied and crosses threshold value discontinuous density jumps appear and indicate transition from capacitive (E) to inductive (H) mode and inductive(H) to helicon (W) mode\cite{degeling1996plasma}. Helicon mode is also known as wave mode coupling since power couples through the wave. So the excitation of the helicon wave is necessary to reach the helicon mode of discharge and the condition for helicon wave excitation can be seen from the dispersion relation of helicon wave. Which is the dispersion relation of whistler wave in a bounded plasma and with frequency $\omega << \omega_{ce}$, the dispersion relation is given by \cite{chen1996physics},
\begin{equation}\label{eqn:disper1}
  \frac{3.83}{a}=\frac{\omega}{k_z} \frac{e \mu_0 n_0}{B} 
\end{equation}
\begin{equation}\label{eqn:disper2}
   n_0 \propto B 
\end{equation}

where $\omega $ is the angular frequency, $n_0$  the plasma density, $\mu_0$ the permeability, $e$  the electron charge, $k_z$ the axial wavenumber, and $B_0$ the axial magnetic field. From the dispersion relation, it can be seen that the only internal parameter we need to fulfill to excite the helicon wave is density $n_0$. Which is also called density threshold or critical density and it is proportional to the magnetic field. So the density requirement to excite helicon wave at a high magnetic field is higher than the density requirement at a low magnetic field.

\par
The objective of this work is to achieve helicon mode of discharge in the IShTAR plasma source. To realize that first, argon plasma is produced and characterized for a range of neutral gas pressure, magnetic field, and RF power. The global model of particle and power balance was implemented to validate the experimental results as well as to understand plasma parameter dependency on different operating conditions and optimization of operating conditions has been made at the end.

\section{Experimental Setup}

IShTAR is a linear device and it's consists of two parts (as depicted in figure \ref{fig:cross-section}):  the main vacuum vessel which accommodates the ICRF antenna and plasma source.  All the measurements presented done entirely in the plasma source. The plasma source consisted of a half-turn right-handed helical antenna and five small magnetic coils mounted around the glass tube. These coils create a uniform axial magnetic field and help the helicon waves to propagate into the plasma. The strength of magnetic field B in the center of the vessel can be set in the range from 0–64 mT. The helical antenna is 0.64 m long and 0.22 m in radius. It can launch helicon waves with mode m = +1 anti-parallel to B direction and with mode m = -1 waves parallel to B. The antenna is connected to a power generator able to deliver RF power up to 3 kW at around 12 MHz. For the measurements, an RF-compensated single Langmuir probe has been used. The probe is connected to the probe manipulator through an axial shaft to place the probe at a required axial position.
\begin{figure}[htb]
        \centerline{\includegraphics[width=.75\textwidth]{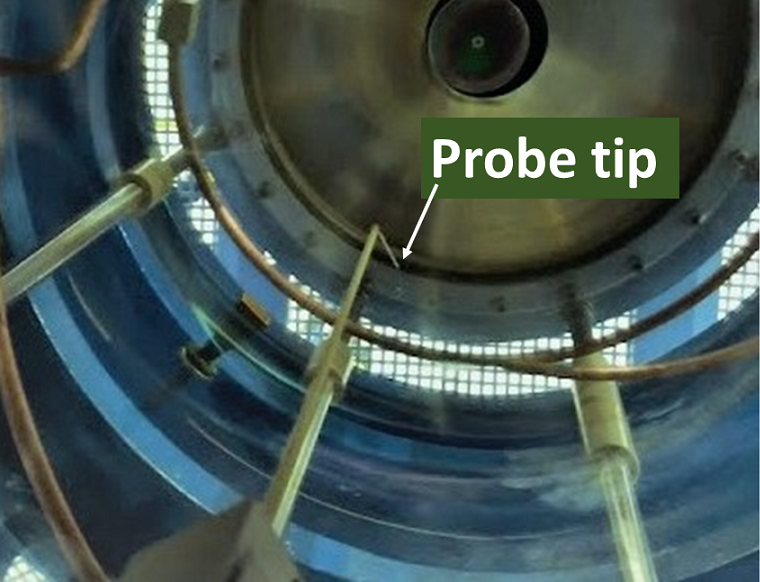} }
        \caption[]{RF-compensated Langmuir probe inside the source chamber}
        \label{fig:probe}
\end{figure}
The probe manipulator system enabling an one-dimensional radial scan with a spatial precision of $\approx$ 2 mm. During the measurement, the radial positioning step size was $\approx 6- 10$mm. In order to resolve better localization of plasma potential features a finer step size has been used.

The probe pin is made from tungsten wire with a length of 15 mm and 0.8 mm in diameter. The probe pin is aligned perpendicular to the axial magnetic field as well as to the direction of the radial scan. Figure (\ref{fig:probe}) showing the Langmuir probe inside the source chamber

% ____________________________________________________________________________________

\section{Plasma characterization} 
\begin{figure}[htb]
        \centerline{\includegraphics[width=1\textwidth]{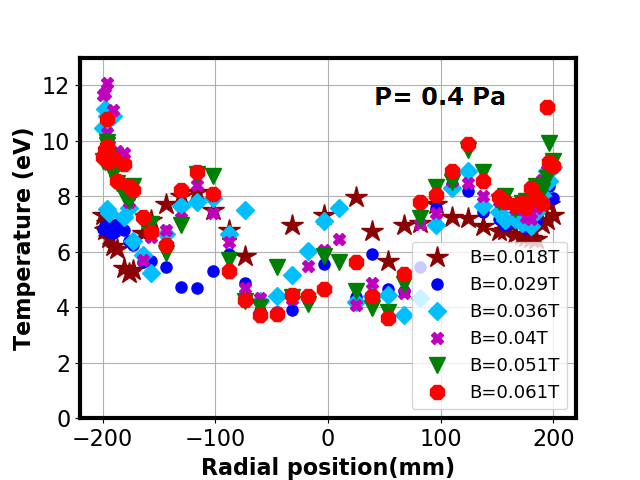} }
        \caption[]{The radial profile of electron temperature at neutral gas pressure, 0.4 Pa and RF power, 3 kW }
        \label{fig:radt}
\end{figure}

\begin{figure}[htb]
        \centerline{\includegraphics[width=.9\textwidth]{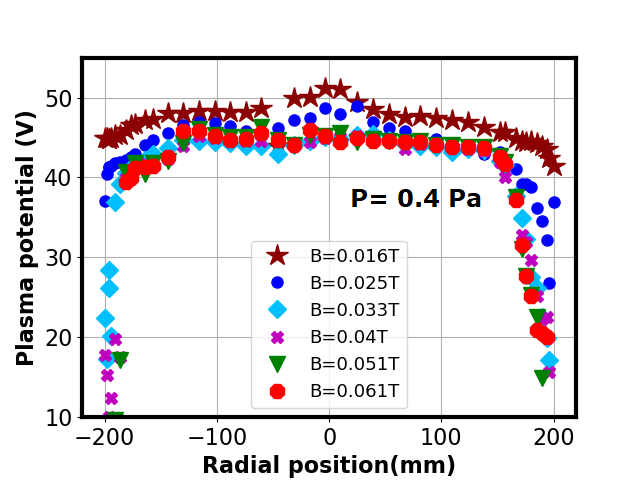} }
        \caption[]{The radial profile of plasma potential at neutral gas pressure , 0.4 Pa and RF power, 3 kW }
        \label{fig:radp}
\end{figure}

\begin{figure}[htb]
        \centerline{\includegraphics[width=1\textwidth]{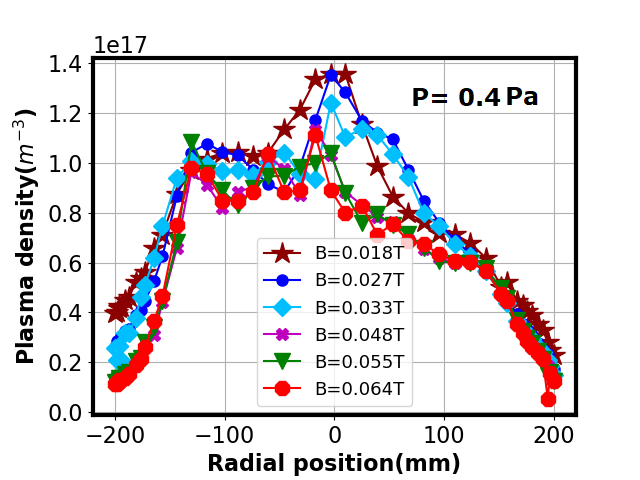} }
        \caption[]{The radial profile of plasma density at neutral gass pressure , 0.4 Pa and RF power, 3 kW }
        \label{fig:radn}
\end{figure}

To optimize the plasma source for the purpose of achieving helicon mode of discharge it is necessary to know the plasma characteristic's in its current configuration. The plasma source radially scanned with the RF compensated single Langmuir probe at different gas pressure in the range from 0.8 Pa to 1.1 Pa and different input RF power in the range from 600W to 3kW. The magnetic field always varies from 0 to 64 mT. Plasma parameters (electron temperature, density, plasma potential) variation with respect to RF power, gas pressure, and magnetic field.  Figure (\ref{fig:radt}), (\ref{fig:radp}), (\ref{fig:radn}) shows radial profile of electron temperature, plasma potential, and density at different magnetic field with a neutral gas pressure 0.4 Pa and input RF power 3 kW. From the radial profile of electron temperature, one can see bulk plasma temperature is lower than the sheath temperature. This is due to the high plasma density in the bulk plasma. The electron temperature dependence on magnetic fields seems insignificant.

\begin{figure}[htb]
        \centerline{\includegraphics[width=.9\textwidth]{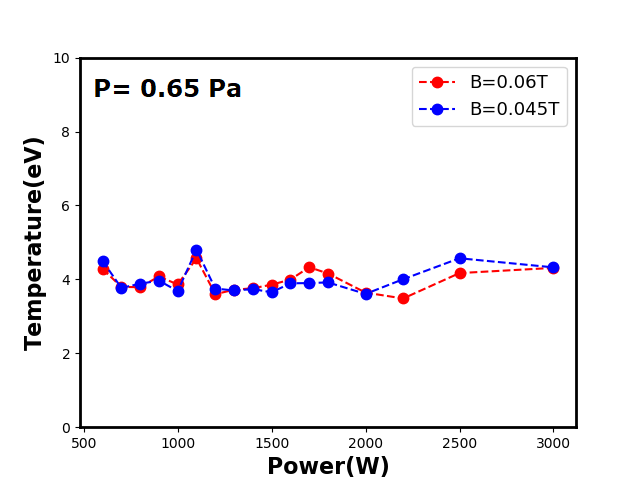} }
        \caption[]{Variation of the electron temperature  with respect to applied RF power and at neutral gas pressure, 0.65 Pa  }
        \label{fig:powt}
\end{figure}

The electron temperature doesn't show any dependence on input RF power as well which can be seen in the figure(\ref{fig:powt}) but temperature shows high dependency on the neutral gas pressure which we can see in the figure (\ref{fig:TeP}). During the experiment, the average electron temperature in the bulk plasma was found to be varied between 4 to 10 eV with the variation of neutral gas pressure.\par
\begin{figure}[htb]
        \centerline{\includegraphics[width=1\textwidth]{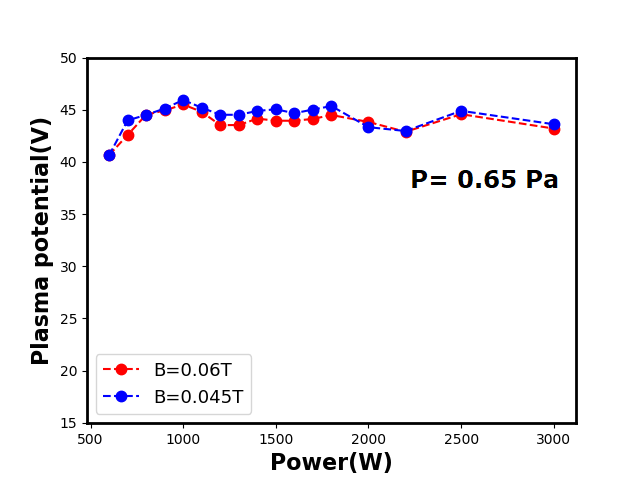} }
        \caption[]{Variation of the plasma potential with respect to applied RF power and at neutral gas pressure, 0.65 Pa  }
        \label{fig:powp}
\end{figure}

Figure (\ref{fig:radp}) shows radial distribution of the plasma potential is uniform and mostly symmetric along the center axis of the source. The plasma potential dependence on input RF power is negligible which can be seen in the figure(\ref{fig:powp}). The plasma potential shows little dependence on magnetic field as well as on neutral gas pressure. The plasma potential varies between 40 to 50 V depending on the magnetic field and neutral gas pressure. \par         

The plasma density shows a strong dependency on neutral gas pressure, input RF power as well as on the magnetic field. At neutral gas pressure 0.4 Pa, center-peaked radial density profile observed at the different magnetic field which can be seen in the figure (\ref{fig:radn}). The plasma density dependence on neutral gas pressure and the input RF power can be seen in figure (\ref{fig:NP}) and  (\ref{fig:powerdensxr}) respectively. one can see, plasma density increases with the increase of neutral gas pressure and input RF power. During the experiment, the maximum observed density is, $\approx 2\times10^{17}$ $m^{-3}$ .    

\section{global model}
 The global model is a useful tool to understand plasma behaviors in a low-pressure high-density plasma discharge. This model is based on simple physical laws and can predict spatially averaged quantities like plasma density, plasma potential, and electron temperature. The main objectives to implement the global model in this work is to validate the experimental results and to explain plasma parameters dependency on different operating conditions.\par 
The global model implemented in this work is developed by Lieberman and Lichtenberg\cite{lieberman1994principles}. It consists of two balance equation: particle and power balance. Particle balance predicts electron temperature and plasma potential while power balance predicts plasma density. 
\subsection{Particle balance}

The  Particle  balance  equates  the  total  number  of charged particle  created by ionization in a volume to the total number of particle lost by ambipolar diffusion to the walls. In the ionization process electron-ion pairs created in the bulk plasma mostly due to electron-neutral collisions. The particle loss to the walls mostly ion dominated since electrons are confined by sheaths. Equating creation rate of electron-ion pairs in bulk of the plasma to the rate of particle loss to the walls gives particle balance equation as\cite{lieberman1994principles}, 
\begin{equation}\label{eqn:particle_b}
  K_{iz}(T_e) n_g n_0 \pi R^2l =  u_B n_0 2\pi R(R h_l+l h_R)
\end{equation}

 $K_{iz}$ is the ionization rate coefficient, $n_0$ is the bulk or centre plasma density, $n_g$ is the argon gas density, $u_B=\sqrt{\frac{kT_e}{m_i}}$ is the Bohm velocity at which ions lost to the surface or at the sheath edge, $R$ and $l$ is the radius and length of the plasma source. The parameters $h_R$ and $h_l$ introduce spatial variation into global model. They defined as ratios of plasma density at the center to its value at the sheath edge in the  radial direction, $h_R=\frac{n_sR}{n_0}$ and in the axial direction, $h_l=\frac{n_sl}{n_0}$. Since our diagnostic do radial scan and can measure the density at the sheath edge with good precision so the $h_R$ value in this model calculated by taking center density and the sheath edge density from the experiment. For the axial profile analytical estimation of $h_l$ value used which is given by \cite{lieberman1994principles},
\begin{equation}\label{eqn:hl}
    h_l\approx 0.86(3+\frac{l}{2\lambda_i})^{-\frac{1}{2}}
\end{equation}
Here, $ \lambda_i=\frac{1}{n_g\sigma_i}$ is the ion mean free path and $\sigma_i$ is the ion-neutral collisional cross-section. \par

 Now equation  (\ref{eqn:particle_b}) can be solved for electron temperature, $T_e$. One can see in the equation (\ref{eqn:particle_b}) plasma density $n_0$ cancels out from the equation (\ref{eqn:particle_b})  and as a result electron temperature is independent of plasma density and input RF power. Temperature independence of input RF power can be seen in the figure (\ref{fig:powt}) as well. Once the electron temperature,$T_e$ is obtained, plasma potential can be obtained as a function of, $T_e$. Since the endplate of the source tube is grounded, and if it is assumed net flow of current to the endplate is zero, then the plasma potential can be estimated as,
\begin{equation}\label{eqn:plasmapot}
\Phi_p= T_e [ \frac{1}{2}+ln(\frac{M}{2\pi m})^{\frac{1}{2}} ]
\end{equation}
\begin{figure}[htb]
        \centerline{\includegraphics[width=1\textwidth]{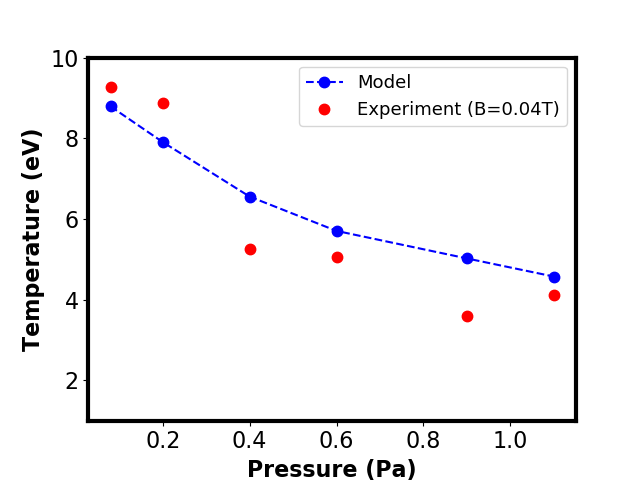}  }
        \caption[]{Electron temperature as a function of neutral gas pressure with a input power of 3 kW.  }
        \label{fig:TeP}
\end{figure}

In the figure (\ref{fig:TeP}), the model predicted electron temperatures are compared with the measured electron temperature as a function of neutral gas pressure at the magnetic field 40 mT. The experimental temperature value used here is the mean temperatures in the bulk of the plasma. One can see a qualitative agreement between model-predicted results and experimental results. Though there is a deviation in experimental value from the predicted model, the trends are similar. The electron temperature decreases with the increase of neutral gas pressure. This is due to the fact that electron temperature is inversely proportional to the neutral gas density. 

\par
\begin{figure}[htb]
        \centerline{\includegraphics[width=1\textwidth]{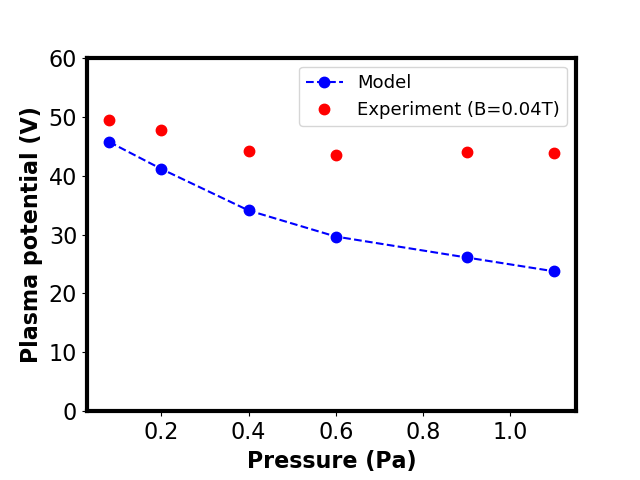}  }
        \caption[]{Plasma potential as a function of neutral gas pressure with an input power of  3 kW. }
        \label{fig:PP}
\end{figure}
However, measured plasma potential shows a high deviation from the model predicted results at higher neutral gas pressure which can be seen in the figure (\ref{fig:PP}). The exact reason for this deviation is not well understood. One possible reason could be the existence of non-ambipolar diffusion and the appearance of the short-circuit effect\cite{simon1955ambipolar} at the conducting endplate at higher pressure. It was found that for gas pressure greater than 0.4 Pa, the ion mean free path, $\lambda_i$  become greater than ion Larmor radius, $\rho_i$. This means ions are non-magnetized at gas pressure greater than 0.4 Pa and the diffusion is no longer ambipolar. In such a situation, particle flux across the magnetic field is ion dominated while the particle flux along the magnetic field is electron dominated. The diffusion becomes non-ambipolar and flux tubes become short-circuited at the conducting endplate. This results in the flow of current through the plasma and the metallic endplate. Whereas in the model we considered ambipolar diffusion and no current flowing to the endplate. In the figure(\ref{fig:PP}) one can see there is a similar decreasing trend of plasma potential between model and experiment till the gas pressure 0.4 Pa, where diffusion is ambipolar.

\subsection{Power balance}
 Power  balance equates RF power absorbed in the plasma, $P_{abs}$, to the power loss because of the energy carried by the plasma leaving ions and electrons. From power balance, we can have the bulk or center plasma density, $n_0$, which is given by,
 \begin{equation}\label{eqn:pn}
n_0=\frac{P_{abs}}{q  u_B A_{eff} E_T(T_e)}
       \end{equation}
Here, $A_{eff}$ is the effective particle loss area defined by,  $A_{eff}=2\pi R(Rh_l+lh_R)$. $E_T(T_e)$ is the total energy loss per ion-electron pair which includes loss due to ion-neutral collisions, electron-neutral collisions, and the kinetic energy loss of electrons, ions species during recombination at the glass tube boundaries. After knowing the temperature, $T_e$, from the equation (\ref{eqn:particle_b}) plasma density can be calculated from the equation (\ref{eqn:pn}) as a function of input parameters.\par
\begin{figure}[htb]
        \centerline{\includegraphics[width=1\textwidth]{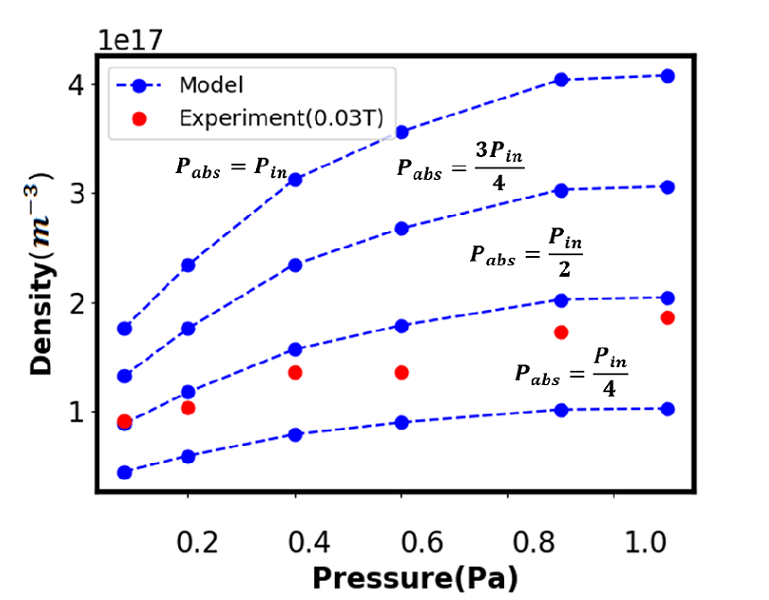}  }
        \caption[]{Plasma  density  in  the bulk plasma  as  a  function  of neutral gas pressure with an input power  $3kW$.}
        \label{fig:NP}
\end{figure}
Due to technical difficulties, it was not possible to measure absorbed power during the experiment but we can have a rough approximation of absorbed power in the plasma from power balance by considering the different percentage of input power as absorbed power and then comparing the model predicted density values with the experimental results. For the model four different percentage of input power considered which are 100, 75, 50 and 25 \% of the input power, $P_{in}$ (i.e. $P_{abs}=P_{in}, \frac{3P_{in}}{4}, \frac{P_{in}}{2}, \frac{P_{in}}{4}$ ). Figure (\ref{fig:NP}) shows plasma density as a function of neutral gas pressure. The plasma density increases with the increase of gas pressure and the experimental result is qualitatively in agreement with the model. In this figure, one can see the absorbed power in the plasma in between 25 to 50\% of the input power.\par  

\begin{figure}[h!]
  \begin{subfigure} [b]{1\textwidth}
    \includegraphics[width=\textwidth]{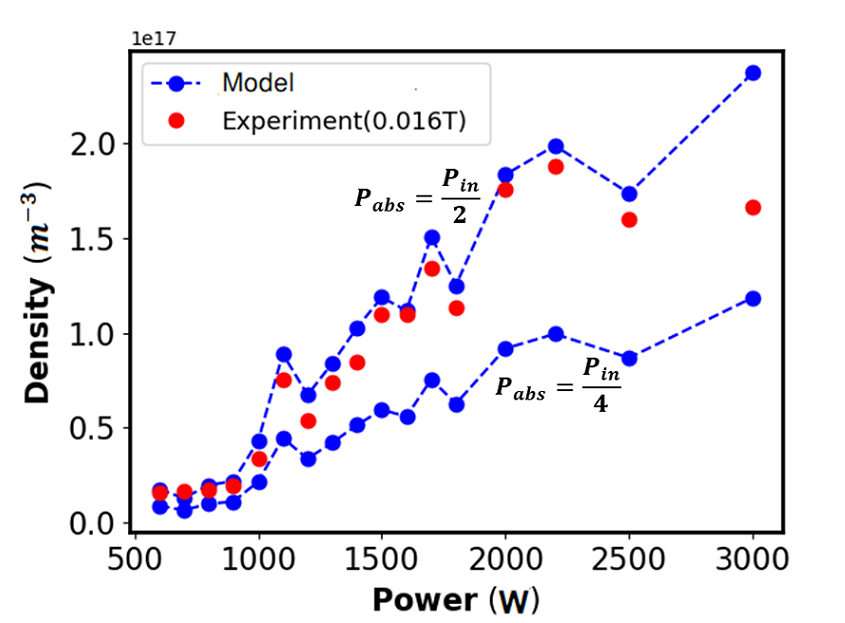} 
    \caption{At magnetic field 16 mT}
    \label{fig:podens7}
  \end{subfigure}
  \hfill
  \begin{subfigure}[b]{1\textwidth}
    \includegraphics[width=\textwidth]{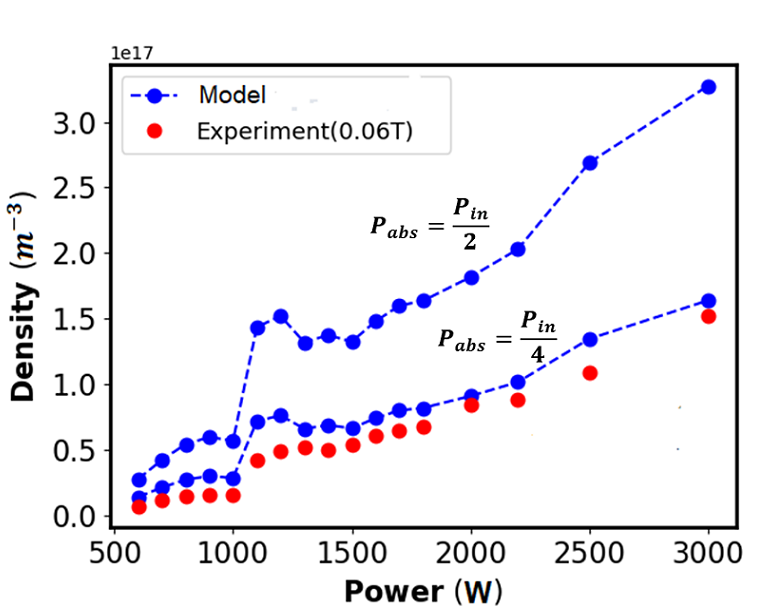}
    \caption{At magnetic field 60 mT}
    \label{fig:podens8}
  \end{subfigure}
  \caption{ Plasma density in the bulk plasma as a function of applied RF power with a neutral gas pressure of $0.65 Pa$. }
  \label{fig:powerdensxr}
\end{figure}

The variation of the plasma density with respect to the input RF power can be seen in the figure (\ref{fig:powerdensxr}). Figure(\ref{fig:podens7}) and (\ref{fig:podens8}) shows density variation at magnetic field 16 mT and 60 mT respectively. One can notice that absorbed power is higher at the low magnetic field than at the high magnetic field. At 16 mT the absorbed power is close to 50\% of the input power while at 60 mT the absorbed power is close to 25\%. This indicates at a low magnetic field high power coupling mechanism possibly exists. This high power coupling mechanism could be the helicon wave mode coupling but that needs to be verified.

\section{Mode transition}
 One way to verify the existence of helicon mode is by observing mode transition in the plasma discharge when the strength of RF input power and magnetic field is varied. One of the basic features of mode transition is that a discontinuous jump in density occurs when plasma discharge transit from capacitive (E) to inductive(H) mode as well as from inductive (H) to helicon mode(W) of discharge\cite{degeling1996plasma}. There is a distinctive difference between the mode transition from capacitive (E) to inductive (H) mode and from inductive (H) to helicon (W) mode. The capacitive (E) to inductive (H) mode transition independent of the applied magnetic field while the inductive (H) to helicon (W) mode transition dependent on the applied magnetic field. For different magnetic fields, the transition to helicon mode occurs at different RF power. This is due to the fact that the critical density needed to excite helicon mode is proportional to the magnetic field which one can see from the dispersion relation (\ref{eqn:disper1}) of helicon wave. So at a low magnetic field, the critical density requirement is less that means low input power is required to achieve that density, and at the high magnetic field, the input power requirement will be higher. \par
\begin{figure}
  \begin{subfigure} [b]{.95\textwidth}
    \includegraphics[width=\textwidth]{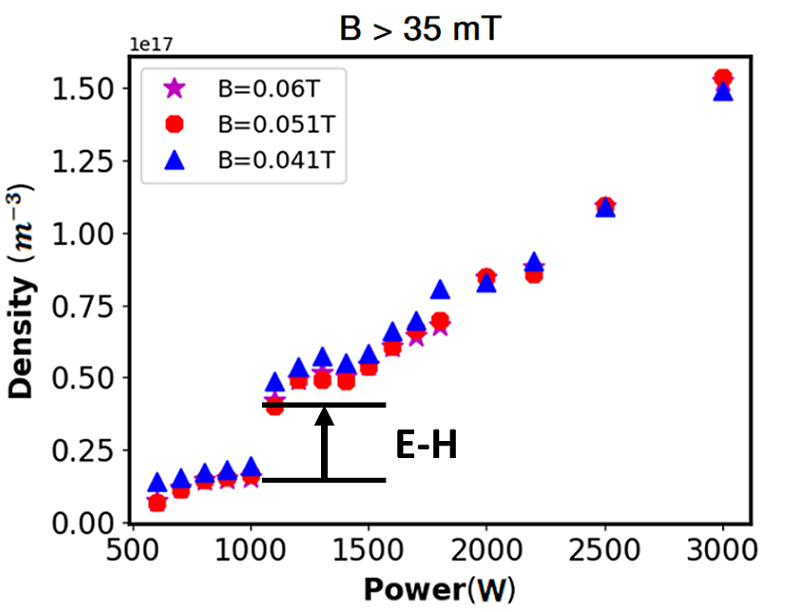} 
    \caption{For magnetic fields, B>35 mT}
    \label{fig:densjump1}
  \end{subfigure}
  \hfill
  \begin{subfigure}[b]{1.01\textwidth}
    \includegraphics[width=\textwidth]{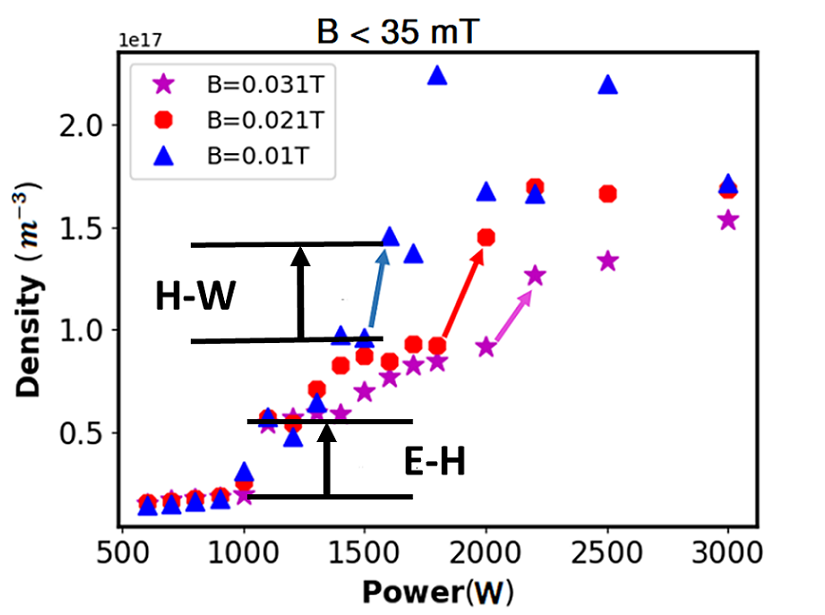}
    \caption{For magnetic  fields, B < 35 mT}
    \label{fig:densjump2}
  \end{subfigure}
  \caption{Plasma density vs power with a working gas pressure 0.65 Pa  }
  \label{fig:denjump}
\end{figure}
Figure (\ref{fig:denjump}) shows two density vs power plot: figure (\ref{fig:densjump1}) represent plot for magnetic field, B > 35 mT (B= 60, 51, 41 mT) and figure (\ref{fig:densjump2}) represent plot for magnetic field, B < 35 mT (B= 31, 21, 10 mT). One can see in both plots first density jump appears around 1000 W and irrespective of magnetic fields the density jump appears at the same power level. So this density jump around 1000 W represents magnetic field independent capacitive (E) to inductive (H ) mode transition. In figure (\ref{fig:densjump1}) no other density jumps can be seen after the first one and density increases in a linear fashion with power. This means discharge does not reach helicon mode at these high magnetic fields. In figure (\ref{fig:densjump2}) one can see,  one more density jumps occur around 1500w, 1800W, and 2000W for the magnetic field 10 mT, 21 mT, and 31 mT respectively. Density jumps at different power levels depending on the magnetic field strength are in agreement with the helicon wave dispersion relation and it confirms that IShTAR plasma source reaching helicon mode of discharge at lower magnetic fields (B < 35 mT). So, the high absorption of power at the low magnetic field which has been seen in the figure (\ref{fig:podens7}) is due to the presence of helicon wave coupling mode. 

\section{Optimization}
Though IShTAR source can reach helicon mode of discharge at low magnetic fields but to achieve more dense plasma IShTAR source need to reach helicon mode of discharge at high magnetic fields (B > 35 mT). So, to find the optimized condition to reach the helicon mode of discharge for the magnetic field, B >35 mT, an analytical study has been done. The threshold density or critical density to excite helicon wave at different magnetic fields calculated from the helicon wave dispersion relation. The dispersion relation (\ref{eqn:disper1}) can rewrite as, 
 \begin{equation}\label{eqn:density}
     n_0=\frac{3.83 k_z B } {a e \mu_0 }
 \end{equation}
 $a$ is the source radius and axial wave number, $k_z$  as a function of antenna length, $l_a$ is given by ,
  \begin{equation}\label{eqn:axialk}
     k_z=(2n+1)\frac{\pi}{l_a}
 \end{equation}
 where $n$ is the longitudinal mode number ($n=0, 1, 2, 3..$). The IShTAR source radius, $a = 0.2 m $ and the antenna length, $l_a = 0.64 m$. The IShTAR power generator can deliver energy with a frequency of up to 15MHz. With varying frequency (6-15 MHz) and magnetic field (0-64mT) a detailed study for critical density has been made. It has been seen for a given magnetic field the critical density requirement is lower at a higher frequency. However, due to antenna matching the frequency is fixed at 12.04 MHz, and at this frequency, the critical density at the highest magnetic field 64mT is $3.9\times 10^{17}m^{-3}$ while the maximum experimental value we achieved at this magnetic field is $1.78\times 10^{17}m^{-3}$ for gas pressure 1.1 Pa with an input power 3 kW.  So, we are not far from achieving helicon mode of discharge at a high magnetic field for the longitudinal mode, n=0. But for the longitudinal mode, n=1 at magnetic field 64mT  the density requirement is higher which is $1.18\times 10^{18}m^{-3}$. On the other hand, from table (1) one can see the density requirement to excite helicon wave at the magnetic field 10 mT is $6.17\times 10^{16}m^{-3}$ whereas during the experiment at this magnetic field the density value achieved is $9.59\times 10^{16}m^{-3}$ at gas pressure 0.65 Pa and 1.5 kW input power. This also proves that we indeed achieving helicon mode of discharge at lower magnetic fields.\par
 
 \begin{table}
\label{table1}
\caption{Comparison between critical density ($n_0$) from dispersion relation and maximum density achieved during experiment at low (10 mT) and high (64 mT) magnetic field  }
\begin{tabular}{| c| c|p{2.7cm}|p{2.7cm}| }
\hline
$n $  & $B$  & Critical density ($n_0$)& Maximum acieved density\\
\hline
0 & 10 mT & $6.17\times 10^{16} m^{-3}$ &\textcolor{red}{ $9.59\times 10^{16} m^{-3}$} ($P$ = 0.65 Pa, $P_{in}$= 1.5 kW)\\
\hline
0 &64 mT & $3.9\times 10^{17} m^{-3}$& $1.78\times 10^{17} m^{-3}$ ($P$ = 1.1 Pa, $P_{in}$= 3 kW) \\
\hline
1 & 64 mT &  $1.8\times 10^{18} m^{-3}$& \\

\hline
\end{tabular}
\end{table}
Since we have a power balance model it is possible to know the approximate amount of power need to be absorbed in the plasma to reach these critical densities. Equation (\ref{eqn:pn}) can rewrite as,
\begin{equation}\label{eqn:pn1}
P_{abs}=n_0q  u_B A_{eff} E_T(T_e)
\end{equation}

Since $T_e$, as well as $h$ factors which included in $A_{eff}$ are a function of gas pressure the $P_{abs}$  also become a function of gas pressure and it has been seen that at higher gas pressure the absorb power requirement is lower to achieve a given critical density. This is also in agreement with other experiment\cite{sharma2018development}. The maximum gas pressure the IShTAR source can have is 1.2 Pa and at this gas pressure the amount of power need to be absorbed to reach critical density at magnetic 64mT, is 1800W. From the figure(\ref{fig:podens8}) one can see the power absorption at a high magnetic field close to 25\% of the input power. By assuming 25\% power absorption before the transition to helicon mode the approximate amount of input power needs to reach helicon mode of discharge at 64 mT is $\approx$ 7.2 kW. For longitudinal mode n = 1, the approximate input power required to reach helicon mode at 64 mT is $\approx$ 20 kW. However, the maximum available power to IShTAR source is only 3 kW. So a high power generator needs to be installed to reach helicon mode at higher magnetic fields (B > 35 mT). 
\section{Summary}
Experiments have been carried out and a discharge model implemented to validate the experimental results. The plasma parameters dependency on the different operating conditions has been investigated. The electron temperature and density were found to be in agreement with the model. The temperature shows a stronger dependency on gas pressure while density has a stronger dependency on gas pressure, Rf input power as well as on the magnetic field. The plasma potential shows a high deviation with the model at higher gas pressure. One possible reason for this deviation could be due to the fact that at higher pressure (P > 0.4 Pa) ions become non-magnetized and diffusion become non-ambipolar while in the model diffusion considered ambipolar. From power balance, it has been seen that power absorption is high at a low magnetic field which indicates that at a low magnetic field an efficient power coupling mechanism possibly exists. The density vs power plot also shows helicon type density jump at low magnetic fields ( B < 35 mT) which substantiates that for magnetic fields, B < 35 mT, IShTAR plasma source can sustain in helicon mode of discharge. However, to achieve more dense plasma IShTAR source needs to reach helicon mode at higher magnetic fields (B > 35 mT) and optimization shows that a high power generator needs to be installed.

%\section{Acknowledgement }

\bibliography{bibliography.bib}
\bibliographystyle{unsrt}
%_________________________________________

%_________________________________________
\newpage

\end{document}